\documentclass[11pt]{scrartcl} 

\usepackage{url,hyperref,lineno,microtype,subcaption}
\usepackage{booktabs}
\usepackage{graphicx}
\usepackage{hyphenat}
\usepackage[
  style=apa,
  backend=biber,
  sorting=nyt
]{biblatex}

\DeclareLanguageMapping{english}{english-apa}

\title{Caring Over Computing: An Ethical and Sociotechnical Perspective on Generative AI for Social Connectedness in Dementia Care}

\begin{document}

\thispagestyle{plain}

\begin{center}
{\Large\bfseries
Caring Over Computing: An Ethical and Sociotechnical Perspective on Generative AI for Social Connectedness in Dementia Care\par}

\vspace{2em}

\begin{tabular}{cc}
\begin{tabular}{c}
\textbf{Teis Arets}$^{1,*}$\\[0.3em]
\small Human Technology Interaction\\
\small Eindhoven University of Technology\\
\small Eindhoven, The Netherlands \\[3em]
\end{tabular}
&
\begin{tabular}{c}
\textbf{Giulia Perugia}$^{1,\dagger}$\\[0.3em]
\small Human Technology Interaction\\
\small Eindhoven University of Technology\\
\small Eindhoven, The Netherlands \\[3em]
\end{tabular}
\\[2em]
\begin{tabular}{c}
\textbf{Maarten Houben}$^{2,\dagger}$\\[0.3em]
\small Department of Industrial Design\\
\small Eindhoven University of Technology\\
\small Eindhoven, The Netherlands
\end{tabular}
&
\begin{tabular}{c}
\textbf{Wijnand IJsselsteijn}$^{1}$\\[0.3em]
\small Human Technology Interaction\\
\small Eindhoven University of Technology\\
\small Eindhoven, The Netherlands
\end{tabular}
\end{tabular}

\vspace{1.5em}

\small
$^*$ Corresponding author:
\href{mailto:t.t.j.e.arets@tue.nl}{t.t.j.e.arets@tue.nl}

$\dagger$ These authors contributed equally.
\end{center}

\begin{abstract}

People with dementia in residential care often experience reduced social connectedness. 
Person-centered care approaches foreground meaningful social interactions to support well-being, but staff and time pressures increasingly constrain opportunities for such care.
Generative AI (GenAI) technologies offer possibilities for helping care professionals facilitate social connectedness, despite increasing care pressures.
Yet, the meaningful integration of such technologies into the complex sociotechnical setting of residential dementia care is not straightforward. 
To responsibly design GenAI technologies for dementia care, we need to investigate how care professionals recognize and foster social connectedness and how technologies intersect with those approaches in care contexts.
We conducted five workshops with 15 care professionals from three care organizations to pursue this objective. 
Our findings show that fostering social connectedness involves ongoing interpretation, negotiation, and adaptation to residents’ individual needs. 
Technology was used by care professionals to support communication, interaction, and shared activities, yet remained largely peripheral to everyday care routines and required active mediation for use in practice. 
Drawing on an ethics of care framework, we derive ethical and sociotechnical implications for the responsible design and positioning of GenAI in residential dementia care.
\end{abstract}


\section{Introduction}

Dementia refers to neurodegenerative conditions characterized by progressive decline in cognition, physical abilities, and psychosocial well-being \parencite{world_health_organization_dementia_2023,world_health_organization_global_2021}, affecting over 55 million people worldwide.
Dementia directly impacts social health, reducing opportunities for social interaction, diminishing the ability to participate in social activities \parencite{chung_activity_2004}, and increasing the difficulty of following and sustaining everyday interactions \parencite{farrell_subjective_2014,banovic_communication_2018}. 
These challenges limit opportunities to experience social connectedness, the sense of belonging to a social group or network \parencite{heins_effects_2021,van_bel_social_2009}, especially upon admission to residential care \parencite{williams_transitions_2025}.

Person-centered dementia care approaches 
recognize social connectedness as central to the well-being and personhood of people with dementia \parencite{kitwood_dementia_1997,brooker_person-centred_2015}.
In the context of person-centered dementia care, care professionals are positioned as primary actors in fostering social connectedness, not only because of their continuous presence, but also due to their role in facilitating everyday interaction and supporting residents' socio-emotional well-being \parencite{garnett_health_2024}.
Yet, the enactment of person-centered dementia care increasingly takes place under conditions of staff shortages \parencite{heger_personnel_2025}, increasing workload and emotional strain among care professionals \parencite{van_aerschot_psychophysical_2022}, and time-constrained care practices within residential care environments \parencite{ludlow_staff_2020,johnson_residential_2010}.
These conditions restrict opportunities for meaningful social interaction and engagement, despite the recognized importance of the latter for the well-being of people with dementia.

In recent years, technologies have increasingly been introduced to support care professionals in fostering social connectedness among people with dementia \parencite{ma_bridging_2024}.
In particular, the emergence of Generative AI (GenAI) models, systems that produce seemingly novel content, such as text, images, or videos, in response to user prompts, drawing on patterns learned from large datasets \parencite{feuerriegel_generative_2024,ooi_potential_2023}, introduces new possibilities for supporting care professionals in fostering social connectedness. 
Through capabilities such as generating conversational prompts, adapting interactions to individual preferences, producing personalized materials, or facilitating open-ended dialogue in real time, GenAI may support care professionals in creating and sustaining meaningful social interaction \parencite{feuerriegel_generative_2024,li_always_2025,maeda_when_2024,koubaa_exploring_2023}. 
For instance, GenAI has been used to support social companionship, reminiscence, and storytelling with people with dementia 
\parencite{xygkou_mindtalker_2024,arets_shared_2026,seah_rememo_2026,zhang_remihaven_2025}.

However, GenAI systems do not operate independently of existing care practices, but, like many other technologies, inevitably reshape routines, interactions, and responsibilities within complex sociotechnical settings \parencite{greenhalgh_beyond_2017,suchman_human-machine_2006}. 
Thus, the introduction of GenAI into dementia care for social connectedness raises not only practical questions about how such systems can be developed for this context, but also ethical and sociotechnical questions about how they can be responsibly positioned within the relational processes through which care professionals recognize and enact social connectedness in practice.
To understand these ethical and sociotechnical implications, it is therefore important to find out how care professionals understand and foster social connectedness in everyday care practice, as well as how technologies are currently positioned within these practices. 
To this end, we pose the following research questions:

\begin{enumerate}
    \item[\textbf{RQ1}] How do care professionals understand and approach fostering social connectedness for people with dementia in residential care?
    \item[\textbf{RQ2}] How might GenAI be responsibly positioned to support social connectedness in residential dementia care?
\end{enumerate}

We draw on Tronto’s (\citeyear{tronto_moral_1993}) ethics of care framework to structure our discussion and reflect on our findings.
This paper makes three contributions:
(1) it describes how care professionals recognize, negotiate, and enact social connectedness in residential care;
(2) it conceptualizes social connectedness as a relational care process through which social needs are continuously interpreted and enacted; and 
(3) it unpacks ethical and sociotechnical considerations for the responsible positioning of GenAI technologies for promoting social connectedness in dementia care.

\section{Background}
\subsection{Social Connectedness in Dementia Care}
Social connectedness 
is associated with factors such as relationship salience, closeness, shared understanding, and quality of social contact \parencite{van_bel_social_2009}.
The experience of social connectedness positively affects quality of life and mental well-being \parencite{kitwood_towards_1992,martyr_living_2021}.
However, people with dementia often experience reduced opportunities for social connectedness, particularly in residential care settings, as cognitive and physical decline limit communication abilities \parencite{alsawy_what_2017,xu_social_2024,banovic_communication_2018} and opportunities to engage in social interaction \parencite{giebel_activities_2015}.
As a result, many people with dementia live with unmet social needs \parencite{moyle_grand_2023,cohen-mansfield_which_2015,orrell_needs_2008}, which are often expressed indirectly through so-called responsive behavior \parencite{moyle_grand_2023,cohen-mansfield_which_2015,alzheimer_society_of_canada_dementia_2019,song_factors_2019}. 
In residential care, care professionals play a central role in interpreting such responsive behavior and translating it into social needs that can be acted upon \parencite{orrell_needs_2008}.
Accordingly, care professionals often rely on familiarity, observation, and knowledge developed through ongoing everyday interactions to recognize social connectedness needs \parencite{novy_relational_2023}.

At the same time, emerging technologies are increasingly introduced in residential dementia care, including socially assistive robots \parencite{shi_exploring_2025}, VR-based applications \parencite{waycott_role_2022}, and other digital technologies supporting care work and resident well-being \parencite{ma_bridging_2024}. 
Within these evolving care contexts, understanding how care professionals perceive and foster social connectedness becomes increasingly important.

\subsection{Social Connectedness Through a Lens of Ethics of Care}
To unpack how care professionals perceive and foster social connectedness, we draw on ethics of care, a relational sociotechnical perspective on caregiving that positions care as grounded in relationships, responsibility, and situated responses to the needs of others, rather than abstract moral rules \parencite{gilligan_different_1993,waycott_role_2022,haraway_modest_witnesssecond_millennium_1997}.
We therefore conceptualize fostering social connectedness as a relational care practice through an ethics of care lens.

To operationalize this perspective, we draw on Tronto’s (\citeyear{tronto_moral_1993}) framework of care, which conceptualizes care as a process comprising distinct but interrelated phases, each associated with specific moral qualities. 
This framework allows us to analyze how social connectedness is fostered as a situated care practice.
The first phase, \textit{caring about}, involves noticing a need for care \parencite{tronto_moral_1993}.
In the context of this study, this phase involves the care professional's recognition of social needs in people with dementia.
The second phase, \textit{taking care of}, requires the care professional to assume responsibility for the identified social needs and determine how to respond to them \parencite{tronto_moral_1993}.
The third phase, \textit{care giving}, concerns the enactment of care \parencite{tronto_moral_1993}. 
This step involves carrying out that intention, for instance, by engaging in a meaningful interaction with a person with dementia or by having a loved one stop by.
The fourth step, \textit{care receiving}, shifts attention to the care recipient and involves whether the need for care---or, in our case, the social need---has actually been met.
An important caveat to this framework is that in practice, the phases of care are never as linear as they appear here \parencite{tronto_moral_1993}.
Nevertheless, this ethics of care framework helps us make the practices that foster social connectedness visible,
evaluate whether and how emerging technologies like GenAI can assist in fostering social connectedness, and position them within the relational processes through which social needs are recognized and addressed in residential dementia care.

\subsection{GenAI for Social Connectedness}
GenAI refers to AI-based systems that generate seemingly new content based on a user's request (i.e., prompt), using large datasets to compose the output most likely to align with such a request \parencite{feuerriegel_generative_2024,ooi_potential_2023}.
The immediate generative nature of GenAI has potential for supporting the facilitation of social connectedness in residential dementia care, for instance, by recognizing patterns from behavioral data that might indicate a social need \parencite{feuerriegel_generative_2024}, offering decision-support for how to socially engage a person with dementia \parencite{andargoli_intelligent_2024,fadul_generative_2025}, or acting as an LLM-based interaction partner itself \parencite[e.g.,][]{hossain_chatgpt_2024,shimizu_exploring_2024,xygkou_mindtalker_2024,zhang_reconfiguring_2023}.

A growing line of work taps into the reasoning that people with dementia will benefit from using GenAI to support offline interactions with others, as they see little personal benefit in using conversational agents as social companions \parencite{kot_exploring_2026}.
For instance, GenAI has been used to support collective reminiscence \parencite[e.g.,][]{seah_rememo_2026}, and prompt and output co-creation \parencite[e.g.,][]{arets_shared_2026}.
This line of work foregrounds the relational and interpretive nature of GenAI-supported social interaction rather than treating it primarily as a tool for classification and personalized output.

The introduction of GenAI into residential care with a social purpose inevitably introduces a sociotechnical disruption that reconfigures care practice by reshaping routines, interactions, actors, and responsibilities in everyday care work \parencite{greenhalgh_beyond_2017,suchman_human-machine_2006,latour_reassembling_2005}.
Care professionals play a key role in mediating whether and how such technologies are taken up in practice, as their situated care practices ultimately shape both the use and non-use of technology in residential care settings \parencite{waycott_role_2022,kelly_staff_2025,rai_digital_2022}.
However, this central role also implies that the integration of GenAI into residential dementia care is contingent on how care professionals make sense of its relevance to their everyday practice. 

Despite emerging work exploring GenAI for social interaction \parencite{seah_rememo_2026,arets_shared_2026}, little is known about how care professionals understand and position such technologies within the relational processes through which social connectedness is recognized and enacted. 
One line of work has explored person-centered practices in dementia care from the perspective of care professionals \parencite[e.g.,][]{kitwood_dementia_1997,lee_experiences_2023,novy_relational_2023}, while another has examined how technologies may support care professionals in practice \parencite[e.g.,][]{ma_bridging_2024,kelly_staff_2025,rai_digital_2022}. 
However, there remains a gap in bringing these lines of research together to examine how technologies, such as GenAI-based systems, may fit within care professionals’ ongoing person-centered practices.
This raises critical questions not only about what GenAI \textit{can} do, but about where, how, and whether it \textit{should} be integrated into the process of fostering social connectedness. 
In this paper, we address this gap by investigating how care professionals perceive and enact social connectedness in practice with and without the support of technologies, and position emerging GenAI technologies within this process. 
Drawing on Tronto’s ethics of care framework \parencite{tronto_moral_1993}, we examine how GenAI may intersect with different phases of care and articulate which ethical and sociotechnical boundaries should be enacted for its safe integration in residential dementia care.

\section{Materials and Methods}
\subsection{Study Design}
We conducted five interactive workshops with care professionals to understand their perspectives on the needs for social connectedness of people with dementia living in residential settings, and the use of technological interventions in dementia care. 
The study was approved by the Ethical Review Board of the Human Technology Interaction group at Eindhoven University of Technology (case no. 1978).

\subsection{Participants and Recruitment}
Fifteen care professionals (13 women, 2 men) working with people with dementia, divided into groups of 2 to 4 (see Table \ref{tab:participant_information} for an overview of the group composition), participated in the workshops (between June 2024 and December 2025). 

Participants were recruited digitally by sharing information about the workshops on their care organizations’ intranet, central communication platforms, or via the team leads. The care professionals who participated in the study were employed by three different care organizations (care homes A, B, and C, respectively employing P1--P6, P7--P11, and P12--P15; see Table \ref{tab:participant_information}). Each workshop consisted exclusively of professionals from the same organization. Due to the large size of these organizations, this did not necessarily imply that participants knew each other prior to the study or worked together. The workshops took place \textit{in situ} at each care organization.

\begin{table}[!b]
\centering
\caption{Workshop participants overview. Horizontal lines separate sessions.}
\begin{tabular}{llllp{0.3\columnwidth}}
\toprule
\textbf{Code} & \textbf{Gender} & \textbf{Age} & \textbf{Experience} & \textbf{Profession} \\
\midrule
P1 & F & 30--39 & 3--5 & Certified care assistant\\
P2 & F & 50--59 & 5--10 & Primary care coordinator\\
P3 & F & 40--49 & 20+ & Nurse with coordinating tasks\\
\midrule
P4 & F & 30--39 & 5--10 & Certified care assistant\\
P5 & F & 50--59 & 20+ & Residential support worker\\
P6 & F & 40--49 & 3--5 & Certified care assistant\\
\midrule
P7 & M & 30--39 & 5--10 & Social worker\\
P8 & F & 20--29 & 1--2 & Social worker\\
P9 & F & 50--59 & 10--20 & Certified care assistant\\
\midrule
P10 & F & 40--49 & 20+ & Nurse\\
P11 & F & 30--39 & 5--10 & Psychologist\\
\midrule
P12 & F & 40--49 & 20+ & Nurse with coordinating tasks\\
P13 & F & 50--59 & 5--10 & Certified care assistant\\
P14 & F & 50--59 & 10--20 & Nurse\\
P15 & M & 50--59 & 20+ & Advisor care \& technology\\
\bottomrule
\end{tabular}

\label{tab:participant_information}
\end{table}

Most care professionals in our study worked with people in the mid to late stages of dementia. Our sample included a diverse range of professional roles, some familiar to readers worldwide (i.e., psychologist, nurse, social worker, advisor care and technology), others typical of the care context in which the data were collected. The latter will be described as follows.
\textit{Certified care assistants} provide daily hands-on support to people with dementia in residential care, delivering personal care and basic nursing assistance.
\textit{Primary care coordinators} occupy themselves with maintaining ongoing, direct contact with a designated resident in residential care, supporting daily structure and meaningful engagement while serving as the primary liaison with relatives and the multidisciplinary team.
\textit{Residential support workers} support dementia care teams from a unit-wide perspective, coaching staff and setting up meaningful daily activities.

\subsection{Workshop Procedure}
The workshops started with an introduction, during which the moderators explained to participants their rights (e.g., voluntary participation and withdrawal), data handling procedures, and obtained written informed consent, after which they started the audio and video recordings.

The session consisted of three consecutive phases, each supported by workshop materials which can be found in the Supplementary Materials.
In the \textit{first} phase, participants were presented a table with six statements on the social connectedness of people with dementia.
They first individually reflected on whether they agreed or disagreed with the statements in the table, making notes on post-its, then discussed their reflections with the group.
In the \textit{second} phase, participants used a poster with four quadrants: (1) benefits of social interactions, (2) challenges of social interactions, (3) conversation starters, and (4) activities with a social component. They collaboratively explored each quadrant by adding post-its to the poster and engaging in a group discussion on each other's experiences and insights.
In the \textit{third} phase, participants employed an empty template to map a day in their life and in the life of a person with dementia. 
They used post-its to describe daily tasks and activities, creating a situated representation of how care unfolds in practice, and reflecting on opportunities for social stimulation. 

As a closing discussion topic, participants reflected on their experiences around technology use in dementia care, and provided their perspectives on how technologies are currently used in relation to social connectedness.
After the workshop, participants received a gift worth €12.

\subsubsection{Workshop Setting}
On average, the workshops lasted 1 hour and 35 minutes (range: 1 hour and 27 minutes to 1 hour and 48 minutes). 
The language of the workshops was Dutch.
Each workshop took place in a room with a table large enough to seat five participants and to accommodate the A2-sized workshop materials.

During workshops 1-4, there were two moderators present: one lead moderator and one supporting moderator. The lead moderator guided the exercises and discussions; the supporting moderator took care of distributing and collecting the workshop materials during each phase, took pictures, arranged refreshments, and acted as backup during the discussions. 
During workshop 5, there was only one moderator present.
The first author was the lead moderator in each session; the supporting moderator was an external researcher.

\subsection{Data Collection and Analysis}
We made audio (Voice Memos app on iOS) and video (GoPro) recordings to enable the production of verbatim transcripts annotated with relevant non-verbal behaviors.
The transcripts were generated in the native language of the care professionals with a locally run version of Whisper \parencite{schmidt_audioscribe_2025}, manually corrected while watching the recordings, and anonymized by replacing participant names with the codes listed in Table \ref{tab:participant_information}.

For data analysis, we conducted an iterative, reflexive thematic analysis \parencite{braun_reflecting_2019} of the workshop transcripts to identify patterns within our data relevant to our research questions.
We carried out our analysis following the six-phase framework of thematic analysis originally proposed by \textcite{braun_using_2006}, while also integrating subsequent refinements and methodological guidance \parencite{braun_thematic_2022}.
We conceptualized our approach as an iterative process involving continuous movement across phases and as a reflexive practice that foregrounds the researcher’s situated and active role in the construction of meaning \parencite{bowman_using_2023}.

The first and second steps involved data familiarization and initial coding, which were closely intertwined and hence reported together.
The first author repeatedly read the transcripts for familiarization while inductively coding the text in English using the qualitative data analysis software MAXQDA \parencite{verbi_software_maxqda_2024}. 
While coding, the first author periodically grouped codes into provisional, descriptive categories to keep an overview of the analysis and manage the volume of data. 

The third, fourth, and fifth stages of the analysis (searching for themes, reviewing themes, and defining and naming themes) were interrelated, iterative, and non-linear \parencite{braun_reflecting_2019}. 
After coding the five transcripts (942 codes in total), the first author used MAXQDA to review and cross-compare the codes to identify recurring meaningful patterns. 
Once the initial round of categorization was concluded, the first author re-evaluated each individual code to assess its fit within the existing structure, and then consulted the other authors to review the emerging thematic structure.
This process was repeated multiple times, continuously reassessing key takeaways and (sub-)themes in light of their coherence and alignment with the research questions, reflecting a gradual analytic refinement toward themes most relevant to the study. 
In the final thematic structure, the highest-level groupings served as themes, while lower-level categories comprised sub-themes and narrative elements.

\section{Results}
Thematic analysis yielded four themes\footnote{Digitalized versions of the filled-out workshop materials can be found on OSF: \url{https://lnk.ua/lnmT9p2MV}}:
(1) care professionals assess social connectedness needs by recognizing behavioral expressions over time;
(2) care professionals negotiate social connectedness at the intersection of care practice, regulations, and personal life; 
(3) care professionals use technology to mediate social connectedness in care practice; and
(4) care professionals implicitly shape technology adoption for social connectedness.

\subsection{Care Professionals Assess Social Connectedness Needs by Recognizing Behavioral Expressions Over Time}
Our data revealed that care professionals perceive social connectedness as important for people with dementia. 
However, they acknowledge that recognizing social needs in people with dementia is not exactly straightforward.
In the earlier stages of dementia, one can rely on the verbal skills of people with dementia to articulate feelings of loneliness or a desire for contact:
\textit{``We do have attendees [at daycare] who can express that [social needs] very clearly themselves, like: `well, I am really here for my social contact, because I feel lonely when I sit at home alone' ''} [P8].
However, in later stages of dementia, increasing cognitive and physical impairments make \textit{``expressing those kinds of needs very difficult [...] when you are already that far along in your process''} [P3]. Besides, there is often a stigma associated with expressing social needs, involving a \textit{``piece of vulnerability of opening up [about loneliness]''} [P8].

Care professionals agreed that social needs often surface in the form of behavioral responses to social over- or under-stimulation: \textit{``We see a lot of misunderstood behavior, [...] and very often it lies in, yes, a social aspect''} [P3]. 
For instance, social overstimulation as a result of abundant group activities can cause responsive behavior in residents:
\textit{``We now have a man, and we strongly suspect autism---so this man is somewhat different in social contact. He was once completely angry. Like: `yes, they keep asking me if I want to come play a game, but I do not like that at all. And why do they keep offering that?' ''} [P11]. 
Care professionals emphasize that \textit{``some people do not have a need for social contact''} [P11] and stress the importance of allowing people with dementia to recover from social demands, as well as withdraw from them: \textit{``Because I simply saw that people stayed at their level much longer because they had their own room [where they could withdraw]''} [P12].

Understanding where the need for social stimulation of each person with dementia sits is a complex task.
Apathy or disengagement can be easily interpreted as a person recovering from overstimulation. However, they might as well be indicators of under-stimulation:
\textit{``It is often said, `that person is overstimulated.' But they are also often under-stimulated, and as a result, what remains is, so to speak, that stoic wandering around, for example''} [P3]. 
While in cases of social under-stimulation, increasing social interaction can be considered helpful, the uncertainty about what is going on with the person with dementia makes care professionals doubtful about what to do, as additional stimuli could both alleviate and exacerbate the situation:
\textit{``I find that a challenge, because you can say, that person is under-stimulated, so they need more contact. But maybe they are overstimulated, and that is not something you can just figure out easily''} [P3].

Care professionals reported needing to \textit{``get under someone's skin''} [P9] to---over time---develop an in-depth understanding of them, enabling them to recognize and interpret responsive behaviors:
\textit{``When you get to know someone better, you also learn to pick up on those signals. And when is someone doing well? When they are in balance, when the social, psychological, physical, spiritual---when everything is in balance, that is what you want most. But that is also the biggest challenge, to figure all of that out.''} [P14].
To illustrate how careful observation allows care professionals to infer a social need, P9 shared: 
\textit{``Yesterday, I had to go to a unit to distribute medication. There is a man there and he had somewhat red eyes. So I went to sit in front of him and said: `why do you have red eyes? Do you always have that?' I said: `or have you been crying?' No answer. I said: `do you feel lonely or are you in pain?' `No, I am not in pain.' Then you find out anyway, right? [Then I draw the conclusion] that at that moment he was lonely and missed his wife.''}
Care professionals reported becoming better at distinguishing between over- and under-stimulation this way, by observing people with dementia and continuously assessing how to approach them: 
\textit{``Do you go in very energetically? Or do you go in very calmly? That is really something you have to sense [...] what does this person need from me right now?''} [P3]. 
They acknowledged that the demeanor of care professionals directly shaped the social atmosphere: 
\textit{``If you project calmness [...], the group will also become calm''} [P7].

\subsection{Care Professionals Negotiate Social Connectedness at the Intersection of Care Practice, Regulations, and Personal Life}
Care professionals often perceived themselves as a primary source of social interaction for people with dementia: \textit{``I think that ultimately those who work on the unit are more like family to the clients than their actual family… for the average client''} [P3].
Efforts to foster social connectedness included organizing shared activities such as going \textit{[grocery] shopping together} [P1, P9], \textit{playing games} [P2, P9], \textit{smoking together} [P1, P3, P12], as well as creating environments that invite interaction, such as a \textit{bus stop} [P3] or an \textit{`old men's bench'} [P3] (i.e., a bench in a central place of the residential care locations where residents can come together). 
In addition, care professionals identified conversational topics to facilitate everyday conversations, such as \textit{holidays} [P2], 
\textit{weather} [P7], and 
\textit{pictures} [P9].

However, addressing social connectedness is positioned in a structurally ambiguous space between formal care obligations, personal moral commitments, and institutional priorities.
Within that space, care professionals continuously negotiate whether, how, and to what extent social connectedness can be prioritized.

\subsubsection{Negotiation Between Task-Oriented Care and Relational Engagement in Fostering Social Connectedness}
Part of the negotiation space lies in whether care professionals adopt a relational or task-oriented approach to social connectedness.
From the accounts of care professionals, we should talk about a continuum between relationality and task-orientation rather than a dichotomy.
The relational approach foregrounded person-centered investment in getting to know the person with dementia and involved an intrinsic motivation to recognize and respond to opportunities for social connectedness.
In this context, some described how opportunities for social connectedness could arise in routine care tasks. 
Moments such as \textit{meal moments} [P12], \textit{administering medication} [P11], and \textit{washing} [P7] were not necessarily experienced as inherently social, but could become so when care professionals deliberately engaged with residents during these activities. 
For instance, P5 reflected that care moments could be used to initiate meaningful conversations based on personal items in a person’s room: \textit{``You can also do that during care, if there is a photo frame on the nightstand, to start a conversation during care''} [P5].
At the same time, practicing a relational approach to social connectedness is not without personal cost, but requires care professionals to continuously negotiate how much emotional involvement they can sustain.
For instance, they could develop strong attachments to residents, making it difficult to disengage after a shift: \textit{``...that client who passed away last week, I find that quite difficult. [...] at the moment that she was lying there alone, then I would actually rather not go home. [...] imagine if she passes away here alone tonight. [...] But yes, I also have a private life.[Laughs briefly and humorlessly.] So you do take that with you when you go home''} [P2].

On the other hand, a more task-oriented approach could help care professionals maintain clearer personal boundaries and manage the emotional demands of care work, while potentially limiting the extent to which routine care moments are approached as opportunities for social connection.
In this context, routine care tasks were not always approached as opportunities for social engagement. Some care professionals focused primarily on the functional completion of care tasks:
\textit{``Well, you definitely have care professionals who are purely task-oriented, right. [...] Those who are really task-focused... providing functional care: washing... That, and then not saying anything. Closing the door again.''} [P10].

Participants emphasized that people with dementia are highly sensitive to such differences in approach. 
As P10 reflected, \textit{``... but they [people with dementia] really feel that. And they also say it: [...] `If that [purely task-oriented] person comes, then I do not need care anymore.' They sense perfectly who they have in front of them. [...] That has everything to do with [social] contact.''}

\subsubsection{Negotiation Between Institutional Priorities and Personal Values in Fostering Social Connectedness}
The second part of the negotiation space involved continuously balancing institutional priorities and personal values.
The main institutional factors that constrained opportunities to build personal connections with the residents were limited staffing and time constraints. As P11 noted: 
\textit{``I think that if you can offer those people a fulfilling day, with a fixed structure and also social connectedness, meaningful activities... you can prevent so much. But... care shifts simply do not have time for that. And it is just not feasible.''}
Furthermore, different forms of residential care shaped the extent to which person-centered approaches to social connectedness could be implemented. 
In particular, small-scale living arrangements offered more opportunities for one-to-one interaction, enabling care professionals to better get to know residents and respond to their social needs compared to larger-scale residential care units:
\textit{``...it depends on whether you are in a nursing unit or whether you have small-scale living [...], like at [location 1; small-scale living]. Of course, they are all people with some form of dementia, but at [location 1], people still walked to [supermarket] themselves. [...] [Whereas] when I worked at [location 2; nursing unit], we had a lot of behavioral issues, and that was purely because many people were sitting together in a small living room with eight people, from early morning until late at night. But they were also surrounded by six people whom, in their normal lives, they would probably have avoided by a wide margin''} [P15].

Institutional priorities surrounding safety, health, and risk management further shaped how care professionals navigated opportunities for social connectedness. 
In some cases, care professionals described deliberately working around or flexibly interpreting organizational policies when these were perceived to constrain socially meaningful interactions for people with dementia.
P12 described facilitating an outdoor setting where residents could smoke together: \textit{``A woman does crossword puzzles outside, but she also smokes. Another resident does as well. So they have very nice social contact, with a cigarette outside. Only it is not allowed to be facilitated [...]. But then we thought of maybe a heated blanket. And then we frame it as social contact.''}
In this way, care professionals sometimes prioritized opportunities for meaningful social interaction over medical advice intended to protect physical health (i.e., not smoking).

While some tensions could be navigated through flexible interpretations of institutional policies, participants also described situations involving social connectedness that raised ambiguous moral and professional dilemmas without clear procedural solutions.
For instance, care professionals described situations in which intimacy emerged between residents, requiring careful deliberation about how to respond:
`P15: \textit{``Or two residents who get into bed together?''} 
[Everyone laughs] 
P12: \textit{``Yes, we are actually seriously discussing that proposition right now.''}'
 Elaborating on such a situation, P15 described encountering two residents in an intimate setting during an evening shift:
\textit{``And then he was sitting in that room with his cane [...]. And there was a lady who found him... extremely interesting. [...] Then we had lost track of both of them. I thought oh... I walked in and you really saw there, seriously...''} [P15].

Such situations positioned care professionals in dilemmas where they had to balance respect for residents’ autonomy and intimacy with their responsibility to monitor and safeguard well-being. 
As P12 reflected: \textit{``I was standing at the door [...] what do you do then? Because the moment I walk in... I am no longer respecting their intimacy. [...] But to what extent are you allowed to infringe on that?''} 
In addition, decisions about whether and how to document such situations introduced further ambiguity: 
\textit{``Then, based on your own norms and values, do you report it? And what do you report then? You actually have to report it, because otherwise I cannot do anything with it during the behavioral consultation or the physician's consultation
''} [P12].


\subsection{Care Professionals Use Technology to Mediate Social Connectedness in Care Practice}
Care professionals described a broad range of technologies used in residential dementia care, although only some were explicitly associated with fostering social connectedness.
We identified three configurations through which care professionals leveraged technology: 
(1) to facilitate online communication with loved ones, 
(2) to position technology as the interaction partner, and 
(3) to facilitate a shared medium for offline interaction between people with dementia and others.
Across these configurations, though, care professionals consistently emphasized that technology should be viewed as a complement to, rather than a replacement for human care. 
\textit{``I do think that human contact will always be necessary to make people’s needs known. And perhaps to be able to put it into practice. Because otherwise it could become very cold and impersonal.''} [P7]. 

\subsubsection{Technology for Connecting with Loved Ones}
Care professionals repeatedly mentioned their role in using technology to set up an online communication channel between people with dementia and their loved ones at home, thereby allowing for a sense of social connectedness to emerge.
They did this, for instance, through video calling: 
\textit{``The way they, for example, seek contact with other family members, through video calling. We have quite a few women, especially on the psychogeriatric unit, who are really quite active with their iPad''} [P10]. 
Beyond direct communication, care professionals also envisioned technologies that could help people with dementia maintain a sense of ongoing presence and social connectedness with loved ones, for instance through continuously sharing photos or updates with relatives:
\textit{``I would really like many people to get a digital photo frame. You can just send a photo every time''} [P9]. 

\subsubsection{Technology as an Interaction Partner} 
In a few other cases, care professionals mentioned having tried out technologies that took an active role as an interaction partner of people with dementia, for instance a social robot that initiated engagement: 
\textit{``We already have a robot that indeed says... `Hello, shall we play a game?' `Okay.' And it initiates that''} [P7]. 
However, such technologies were not yet widely embedded in everyday care, with participants suggesting that they might become more relevant for future generations of people with dementia who have had greater exposure to technology throughout their lives:
\textit{``Yes, I immediately pictured such a robot driving around. You also have restaurants where they already... It might be more for the later [upcoming] generations''} [P5].

While social robots had occasionally been introduced into care practice, participants also reflected on emerging forms of conversational AI as potential future interaction partners for people with dementia.
P7 described their own experience with ChatGPT as resembling interaction with another person, and speculated how such technology could provide a sense of meaningful contact in situations of loneliness associated with dementia:
\textit{``I discovered ChatGPT, where I really feel like I am communicating with a person. [...] It feels like personal contact. [...] Suppose I am lonely [...] and I simply get contact... then I would not see a problem with that''} [P7].

\subsubsection{Technology as a Shared Medium for Social Interaction} 
Finally, care professionals described technologies that functioned as a shared medium for in-person social interaction between people with dementia and others (e.g., care professionals, fellow residents, or loved ones), ranging from everyday activities such as listening to music or watching television together to more interactive systems.
For instance, care professionals used an interactive table that could be used for \textit{``playing games, [...] [playing] from the 50s, 60s, 70s, 80s, [...] [practicing] sayings... proverbs...''} [P15] together with people with dementia:
\textit{``We also used it a lot [...] for people who, for example, could not leave their apartment for a few days. So that you would just do something one-on-one''} [P15].
Participants particularly valued projection-based interactive systems for their ability to stimulate collective engagement among residents, which led some care organizations to install additional systems in shared living spaces:
\textit{``We recently had an extra [projection-based interactive system] installed in the living room. So that it [gets used] more...''} [P14].

\subsection{Care Professionals Implicitly Shape Tech Adoption for Social Connectedness}
Despite the different ways in which care professionals leveraged technology with a social purpose, the majority of the care professionals mentioned that technology use remained limited and was used reactively to responsive behavior.

\subsubsection{Technology Remains Peripheral in Everyday Care}
\begin{figure}[t]
    \begin{center}
        \includegraphics[width=\columnwidth]{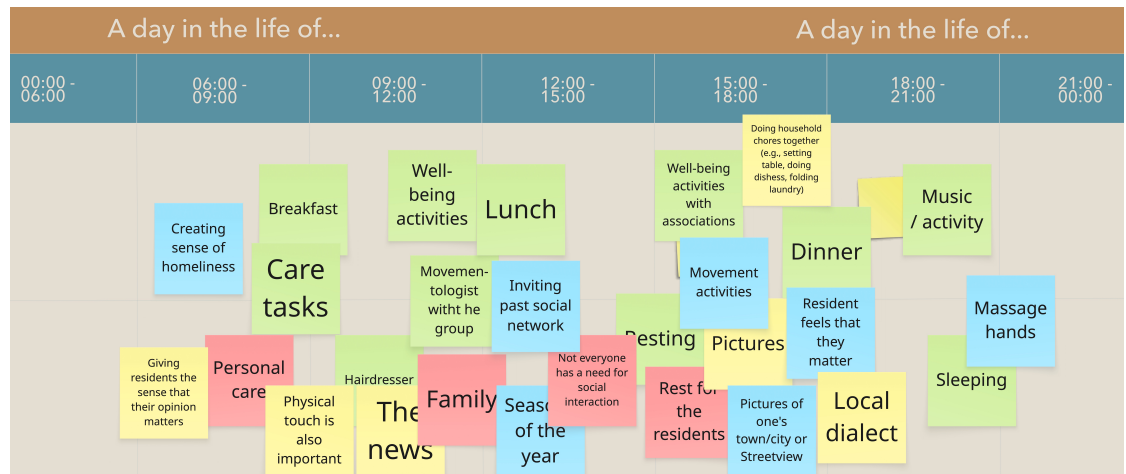}
    \end{center}
    \caption{Digitalized example of schedule mapping by P4--P6. Technology use did not appear in standard routines.}
    \label{fig:post-its_round_3}
\end{figure}

Participants’ accounts of their daily routines revealed that technologies with a social purpose are rarely integrated and often remain unnoticed or underutilized in everyday care, as indicated by the lack of technology-based items on the Post-its in Figure \ref{fig:post-its_round_3}.
Several reasons that partly explain the lack of structural embedding of technologies for social connectedness were mentioned.
First, there was a lack of awareness of the kinds of technologies that are readily available in the care organization:
`P12: \textit{``Yes, but that is it, right. That people often do not even stop to realize that it is there.''}
P13: \textit{``And recently I brought it up, `oh, I did not even know we had that.' ''} 
P14: \textit{``Missed opportunity.''}'
Even when there was awareness of certain types of technologies, their use was not guaranteed. 
Care professionals reported that a lack of knowledge of or confidence to operate certain systems caused technologies for social connectedness to remain unused in practice: 
\textit{``I went to a well-being staff member and I said, `but why are you not using that thing?' I said, `in your role...' [Whispers] `We do not know how it works.' ''} [P15].

Another reason why technologies for social connectedness remained weakly embedded in everyday care practice was that care professionals were not always meaningfully involved in organizational decisions surrounding their introduction:
\textit{``I arrived at the unit there, so I first did an interview, and then that care assistant said, `that interactive table, we do not use it, [...] [because] we did not ask for it! Plus, our team manager thinks, then at least they have such an interactive table. Well, because I was not involved in it, I do not use it!' ''} [P15].
Furthermore, adopting technology often requires deviating from established routines, which can be difficult in practice: 
\textit{``And of course you notice, the longer you have worked in a certain field... `this is how we have always done it,' it also becomes harder to deviate from that''} [P7].
At the same time, participants noted that resistance to new technologies is not fixed, but can shift through exposure and experience, as illustrated by earlier transitions toward digital systems in care: 
\textit{``I remember, in the past we had to switch from paper to computer. [Adopts a dramatic tone] `Oh, that computer...' [...] so much resistance. And after two weeks, everyone was using it.''} [P2].

Participants further described practical and logistical constraints that limited the integration of technologies for social connectedness into everyday care routines. For example, some technologies had fixed locations that required care professionals to move residents or relocate equipment [P2, P13], while others were only available on specific units and therefore required coordination with colleagues before they could be used [P5].
As a result, whether and how such technologies are used largely depends on individual care professionals’ initiative, competence, and ability to integrate them into ongoing care routines.

\subsubsection{Technologies Are Primarily Used Reactively in Response to Observed Behavior}
The peripheral positioning of technologies for social connectedness in everyday care practice shaped how and when they were used. 
Rather than being proactively integrated into ongoing relational care, technologies were typically leveraged in response to observable signs of responsive behavior in people with dementia.
That is, participants reported a desire to deploy technologies when a person with dementia became restless:
\textit{``you just want to be able to seize the moment when someone is restless and then activate something''} [P5]. 

However, their use often depended on individual care professionals recognizing opportunities for meaningful use in the moment.
This became particularly visible in the cases where technologies did become integrated into care practice, often driven by intrinsically motivated `technology ambassadors' who actively promoted and facilitated their use among colleagues.
As P11 commented:
\textit{``And you need a driver, right? So we now have [inaudible] with the SARA robot. But in the beginning, nothing was done with it. It was sitting somewhere in the corner and occasionally, `oh yes, there it is.' But at some point there was an active aging nurse, and she was very engaged with it. `And oh, would it be something for that person, something for that person.' And now it is running. Because now people know, `oh, wait, we have that robot, we can set something up with it.' And then she created a ready-made setting. So the only thing you have to do is place it and press that button.''} [P11].
Intrinsic motivation and enthusiasm for technologies were described as the key to successful adoption. 
If such an ambassador was appointed without that intrinsic motivation, the successful integration of technologies into care practice remained uncertain:
\textit{``We say that we have care technology ambassadors in the organization... but what knowledge are we actually giving those people? None. Are people even intrinsically motivated or... I think a third were simply put forward by the team, like, `well, you just do it' ''} [P15].

\section{Discussion}
This study investigated how care professionals in residential care understand and foster social connectedness in people with dementia (RQ1), as well as how GenAI could fit within these practices (RQ2).
Our findings position social connectedness in dementia care not as a stable state to be detected or delivered, but as a situated relational achievement that emerges through ongoing interpretive care practices.

To interpret our findings and answer RQ2, we adopt Tronto's (\citeyear{tronto_moral_1993}) ethics of care framework and consider how social connectedness is interpreted, negotiated, enacted, and evaluated by care professionals.
For each of the phases of the ethics of care framework, we consider ethical and sociotechnical implications for responsibly introducing GenAI.

\subsection{Caring About}
In terms of noticing a need for social connectedness, our findings show that identifying the social needs of people with dementia increasingly relies on non-verbal communication the more the disease progresses. In line with prior work \parencite{cohen-mansfield_which_2015,theurer_need_2015,moyle_grand_2023,koh_what_2025}, our findings confirm that social needs are typically expressed by people with dementia through responsive behaviors and ascribed by care professionals to social over- and understimulation.
However, our findings further show that translating responsive behavior into social needs is not a straightforward process, but requires a person-centered approach, as unmet social needs do not reside in a specific behavior. 
Rather, the meaning of a particular behavior might differ across individuals with dementia. In residential contexts, this entails that care professionals require profound knowledge of a person's history and behavioral repertoire to infer a social need, and aligns with person-centered care emphasis on seeing people with dementia as individuals with unique histories, preferences, and relational needs rather than through the lens of their cognitive decline \parencite{kitwood_dementia_1997}.

Since social needs recognition appears to be a continuous process of social attunement and situated adjustment \parencite{mol_logic_2008}, we foresee a tension with the prospect of using GenAI to identify or predict needs for social connectedness, for instance, by analyzing behavioral data and/or generating summaries of residents’ states \parencite{bennett_dementia_2025}. 
To infer such needs, GenAI  maps observable signals onto generalized representations learned from training data \parencite{feuerriegel_generative_2024}. However, situated care practice has shown that the recognition of the need for social connectedness is more complex: it requires noticing minute nuances of meaning and situated behavioral changes. 

\textbf{Implication 1}: GenAI systems should not be positioned as means to recognize social needs as they \textit{``necessarily reduce complexity, and (...) remove significant context, in order to make the world more computable''} \parencite[p. 148]{crawford_atlas_2021}. GenAI systems by design filter out the very same contextual and relational nuances through which care professionals come to recognize what a person with dementia needs at a particular moment.
Besides, using GenAI for social needs estimation risks enacting what \textcite[p. 117]{sutherland_alchemy_2019} terms the \textit{``doorman fallacy''}: replacing a role based on its apparent function while overlooking the implicit social and relational value embedded in its performance. 
In this context, the value of recognizing social needs does not lie in its purpose, but in the meaningful human contact that comes with the relational process through which care professionals come to understand a person with dementia. 

\subsection{Taking Care of}
In terms of identifying the right responses to address needs for social connectedness, care professionals described continuously negotiating possible responses based on the preferences of people with dementia, their caregiving styles, available time and staffing, and broader institutional routines and norms. 
Deciding how to foster social connectedness appeared to be a situated, person-centered process, shaped by both relational considerations and the practical realities of residential care \parencite{kitwood_towards_1992,gilligan_different_1993}.
Technologies were rarely described as part of the picture, probably also due to care professionals' unawareness of available technological opportunities, a certain resistance to new ways of working, and the lack of involvement in technology implementation decisions.
Rather, use was described as occasional, tailored to specific situations, rather than routinely integrated into the promotion of social connectedness, and largely dependent on the initiative and literacy of care professionals.
These findings align with prior work showing limited awareness of available technologies in dementia care \parencite{burstein_dementia_2015}, a tendency for interventions to be adopted reactively rather than being embedded within everyday care routines \parencite{kales_assessment_2015}, and a lack of integration of technologies into existing workflows and routines \parencite{iseni_acceptance_2026}. 

In this context, GenAI may enable care professionals to improvise, personalize, and adapt social responses to the preferences, histories, and changing needs of people with dementia.
Unlike traditional technologies that support more predefined activities or interactions \parencite{suchman_human-machine_2006}, GenAI functions as a flexible resource for expanding the possibilities available to care professionals, for instance, by supporting the dynamic exploration of possible responses to situations or by suggesting alternative interpretations and personalized care approaches \parencite{rabbani_generative_2025}.
However, recognizing which possibilities may be meaningful or appropriate in a particular moment for a particular person with dementia should remain a relational process of continuous attunement with the person with dementia, and therefore should be kept in the hands of care professionals.

\textbf{Implication 2}: GenAI may be most ethically appropriate in this phase of care when used to expand care professionals’ possibilities for tailoring and exploring responses without displacing their situated judgment and responsibility.
However, even introducing GenAI as a response support tool comes with a caveat: 
technology use itself may imply new norms, elicit and structure new behavior, and gradually make such technology indispensable. 
This could lead to \textit{``moral tunneling.''}
It may reshape and narrow which social connectedness responses are considered, foregrounding generalized suggestions over more nuanced and context-sensitive possibilities that emerge in practice \parencite{steyvers_three_2024}.

\subsection{Care Giving}
In terms of targeting social needs in care practice, in line with prior work that shows that limited time and staff shift care priorities from meaningful social interaction to fulfilling primary care needs \parencite{ludlow_staff_2020}, our findings show that institutional boundaries often constrain the extent to which social connectedness can be enacted in practice.
In response, some care professionals explicitly use routine care moments as opportunities to promote social connectedness, showing a structural lack of space to specifically and exclusively address social needs in residential care routines. 

Overall, we found that technologies are only rarely embedded in care routines.
Nevertheless, we found that care professionals occasionally used technology to support the enactment of social connectedness in one of three ways: (1) facilitating online communication with loved ones, (2) positioning technology as the interaction partner, and (3) establishing a shared medium for offline interaction with others.
These configurations broadly resonate with the functional triad of technology roles \parencite{fogg_persuasive_2003}, which positions them as tools (i.e., carrying out a task), media (i.e., structuring interactive experiences), or social actors (i.e., stimulating social interaction).
Note that these roles are not mutually exclusive, but their boundaries may blur in practice.
When technologies are used by care professionals, the roles they take are not merely dependent on their functional design, but rather on how they are operationalized in practice.
For instance, a reminiscence tool \parencite[e.g.,][]{xygkou_mindtalker_2024} can be used for individual use, positioning it as a tool (for memory retrieval) and a social actor (for discussing the memory), but also for group use, positioning it as a medium for shared interaction.
Our findings suggest that the social value of these roles does not reside in their epistemological differences, but rather in how they \textit{mediate} meaningful interaction between people with dementia and others, which is determined by how care professionals integrate them into ongoing relational practices.

Requiring such a mediating capacity becomes particularly important in the context of GenAI due to its increasingly human-like conversational capabilities. 
Unlike many earlier technologies used in dementia care, GenAI systems can sustain dialogue, personalize responses, and simulate social attentiveness \parencite{feuerriegel_generative_2024}, thereby more readily positioning themselves as social actors. 
While these characteristics may support engagement, they also create the risk that social connectedness becomes approached as something that can be sufficiently reproduced through human--AI interaction alone, potentially leading to the loss of meaningful contact and relationship-building between people with dementia and care professionals, but also with others. 

\textbf{Implication 3}: GenAI systems should be positioned in ways that allow tool-, medium-, and social actor-like configurations to mediate meaningful interaction between people with dementia and others rather than substitute such interaction. 
Practically, this suggests opportunities for GenAI systems that scaffold shared activities, support co-creation and shared meaning-making, facilitate communication with loved ones, or help care professionals initiate and sustain meaningful interaction around everyday care practices. 
However, their increasingly human-like conversational capacities risk positioning GenAI primarily as autonomous interaction partners, thereby redistributing relational work away from caregivers, loved ones, and peers toward interaction with computational systems themselves. 
In this context, social connectedness risks becoming reduced to an interactional exchange that can be computationally simulated rather than relationally lived \parencite{tronto_moral_1993,mol_logic_2008}.

\subsection{Care Receiving}

Our findings show that, in line with person-centered care \parencite{kitwood_dementia_1997}, care professionals interpret subtle behavioral, emotional, and relational changes to assess whether social needs have been met. Typically,
moments of calmness, reciprocal engagement, participation in interaction, or the disappearance of responsive behavior are indications that social interaction is aligned with the person's needs in that moment.
At the same time, our findings show that social connectedness in dementia care cannot be reduced to maximizing social interaction or stimulation. 
Moments of withdrawal, calmness, or disengagement may not necessarily indicate loneliness or unmet social needs; they can also reflect recovery from social overstimulation. 
As a result, assessing social well-being requires ongoing contextual interpretation of how people with dementia respond to different forms of interaction over time.

This introduces a tension similar to the one identified in the first phase of care. Determining when a need for social connectedness has been met is just as challenging as identifying the presence of such a need in the first place. It is a highly contextual judgment that care professionals make based on subtle behavioral shifts and the situated understanding of what these shifts mean for a specific person with dementia. Such an evaluation cannot be achieved through the mere analysis of behavioral or affective indicators, but requires relational sensitivity and interpersonal attunement.

\textbf{Implication 4}: GenAI may be most ethically appropriate in this phase of care when positioned as a reflective resource that helps care professionals consider how people with dementia respond to different forms of interaction over time, rather than as a system that determines whether social connectedness needs have been met. 
In this role, GenAI could support reflection on patterns of interaction, behavioral change, or relational dynamics without replacing the situated interpretive processes through which care professionals become attuned to the person with dementia. 
Otherwise, GenAI risks narrowing socially meaningful experiences to behavioral or affective indicators that can be computationally interpreted, potentially weakening the relational processes through which social connectedness is understood in practice \parencite{feuerriegel_generative_2024}.

\subsection{Limitations and Future Work}
Our study has several limitations. 
First, because participation was voluntary, our sample likely overrepresents care professionals with greater interest in relational care and technology use. 
Second, the findings are situated within the organizational and cultural context of residential dementia care in The Netherlands, limiting direct transferability to other care systems and contexts. 
Third, our study focused exclusively on residential care, while most people with dementia live at home, where social connectedness is enacted differently and care professionals play a less central role. 
Future research should therefore include more diverse care professionals and care contexts, particularly home care settings involving people with dementia and their social networks. 

\section{Conclusion}
This study explores how care professionals of people with dementia in residential care perceive and enact social connectedness in people with dementia (RQ1) and sheds light on how technologies currently align with those practices, thereby using ethics of care as a moral lens through which potential use of GenAI is evaluated (RQ2).
Findings from workshops with care professionals suggest that recognizing and enacting social needs is a person-centered, relational process at the intersection of personal and institutional values and constraints.
In the context of fostering social connectedness, technology scarcely played a role, stressing that current technologies do not align with the person-centered, relational nature of addressing social connectedness.
Drawing on Tronto's ethics of care framework, we translated the four phases of care to four phases of enacting social connectedness in people with dementia.
Across these phases, our findings suggest that the value of care lies not only in whether social needs are ultimately addressed, but also in the relational processes through which care professionals come to recognize, negotiate, enact, and evaluate those needs in practice.
From this perspective, the introduction of GenAI into dementia care raises ethical and sociotechnical questions that extend beyond concerns about technical performance alone.
While GenAI may support care professionals in fostering social connectedness, our findings suggest that such systems should not replace the relational processes through which social connectedness emerges in practice. Rather, GenAI may be more ethically appropriate when positioned as a mediating resource that supports shared interaction and relationship-building between people with dementia, care professionals, and loved ones. 
In this sense, the ethical implications of GenAI in dementia care concern not only what these systems can do, but also how they may reshape the relational practices through which meaningful social connectedness is understood and enacted.




\section*{Conflict of Interest Statement}
The authors declare that the research was conducted in the absence of any commercial or financial relationships that could be construed as a potential conflict of interest.

\section*{Author Contributions}

\textbf{Teis Arets:} Conceptualization, methodology, validation, formal analysis, writing---original draft, writing---review and editing, visualization, project administration.
\textbf{Giulia Perugia:} Conceptualization, methodology, validation, formal analysis, writing---original draft, writing---review and editing.
\textbf{Maarten Houben:} Conceptualization, methodology, validation, formal analysis, writing---original draft, writing---review and editing.
\textbf{Wijnand IJsselsteijn:} Conceptualization, methodology, validation, formal analysis, writing---original draft, writing---review and editing.

\section*{Funding}
This research was funded by the Dutch Research Council (NWO) through the QoLEAD (Quality of Life by use of Enabling AI in Dementia) Project (project number KICH1.GZ-02.20.008). Additional support from Alzheimer Nederland is gratefully acknowledged. This research was also funded by the research programme Ethics of Socially Disruptive Technologies (ESDiT), under the Gravitation programme of the Dutch Ministry of Education, Culture, and Science and the Netherlands Organization for Scientific Research (NWO grant number 024.004.031).

\section*{Acknowledgments}
We would like to thank Dana Starmans for her support during the workshops. Furthermore, we would like to thank the care organizations Stichting voor Regionale Zorgverlening (SVRZ), Stichting tanteLouise Zorg, and Savant Zorg for hosting the workshops and supporting this research.



\printbibliography





\end{document}